# Low-Code Paradox in DevOps: Security and Governance Insights from Practitioners

Muhammad Azeem Akbar
Software Engineering Department
LUT University
Lappeenranta, Finland
azeem.akbar@lut.fi

Saima Rafi
School of Computing Engineering
& the Built Environment
Edinburgh Napier University
United Kingdom
s.rafi@napier.ac.uk

Arif Ali Khan
M3S Research Unit
University of Oulu
Oulu, Finland
arif.khan@oulu.fi

## ABSTRACT

**Background.** DevOps has become a dominant paradigm in modern software engineering, while low-code development platforms (LCDPs) are increasingly adopted to streamline software development. The integration of these approaches promises efficiency gains but also raises critical concerns regarding security and governance. Despite their growing use, insufficient attention has been given to the implications of these platforms for security and governance in DevOps environment. **Aim.** This study investigates practitioners' perspectives on the security and governance implications of LCDPs in DevOps environments. **Method.** Twelve semi-structured interviews were conducted with IT professionals experienced in low-code and DevOps practices. The data were analysed using a grounded theory approach to identify emergent themes. **Results.** Findings reveal that LCDPs helps in automation of tasks, however, they increase security risks and governance challenges, highlighting the need for robust practices and a security-conscious culture. **Conclusion.** This study suggests that the crossroads of DevOps and LCDPs require careful governance and proactive security practices. Addressing these issues is essential for organisations to unlock the potential of LCDPs while safeguarding resilience, compliance, and the needs for developers.

## 1 Introduction

In today's fast-paced digital landscape, organisations are under increasing pressure to deliver software products on time maintaining quality and compliance. DevOps has emerged as a dominant paradigm in modern software engineering, fostering collaboration between development and operations teams and enhancing the flexibility and efficiency of software development processes [1]. Achieving continuous development, testing, deployment, and delivery, however, requires agility and coordination across all phases of the DevOps lifecycle [2].

Low-code development platforms (LCDPs) have recently gained attention as a means to extend the benefits of DevOps by enabling faster development and deployment of software applications with minimal coding effort. These platforms provide both professional developers and non-technical "citizen developers" with graphical interfaces and automation tools to build applications efficiently [3]. The adoption of LCDPs has accelerated in response to increasing demands for business innovation, labor upskilling, remote work, and economic pressures, with the market expected to reach significant growth in the coming years [4,5]. Despite the valuable assistance, integrating low-code approaches into DevOps environments raises critical concerns regarding security and governance. LCDPs often rely on extensive API integrations and external services, which can expand the attack surface and create challenges related to shadow IT, vendor lock-in, and regulatory adherence [6,7].

This study investigates the crossroadof low-code development and DevOps from the perspective of IT professionals. We explored this research gap by scheduling an interview with twelve practitioners working with LCDPs in DevOps contexts, to explore how these platforms influence security, and governance. Using a grounded theory approach, we analyse the interview data to uncover emergent themes and provide insights into the practical implications of merging low-code and DevOps practices. This study builds on our earlier research [11], the main contribution of this work is to investigate how LCDPs can paradox in DevOps while addressing associated security and governance challenges.

The remainder of the paper is structured as follows: Section 2 presents the background and related work, Section 3 describes the research methodology, Section 4 presents the results, Section 5 discusses and future directions.

## 2 Background and Related work

An approach called DevOps plays a major role in promoting a culture where development and operations teams collaborate closely [1, 2]. It extends Agile principles by encouraging continuous integration, continuous delivery, and iterative cycles, aiming to increase customer satisfaction through rapid updates and services [8, 9]. Originating at the Agile Conference in Toronto, 2008 [10], DevOps has since been widely adopted by major organisations such as Netflix, Facebook, and Yahoo [8, 12] and by cloud-based environment such as Puppet, Docker, Chef, and Amazon services [13, 14]. Researchers have defined DevOps variously as a "cross-functional collaboration between teams" [15], a conceptual framework to support Dev and Ops teams [16], and principles based on Agile and Lean [17]. Its core principles revolve around culture, automation, measurement, and sharing [18], with an emphasis on enabling communication and collaboration [19].

On the other hand, Low-code development platforms (LCDPs) are tools that help in software development by minimising manual coding efforts [7]. Based on graphical user interfaces, templates, and drag-and-drop functionality, low-code platforms allow both professional developers and citizen developers to rapidly build applications [8]. The market for low-code tools is project to reach from USD 37.39 billion in 2025 to USD 264.40 billion by 2032 [20], driven by the need for cost-effective, rapid software development, especially after the COVID-19 pandemic [21]. Recent studies highlight that integrating low-code with DevOps can reduce development complexity, improve time-to-market, and support high-quality software delivery [22]. However, challenges remain in governance, security, and platform fragmentation, as multiple low-code tools may introduce inconsistencies in enterprise environments. Addressing these challenges requires robust governance frameworks and secure DevOps pipelines to fully integrate low-code tools. This research gap motivated us to explore the insights from practitioners' point of view about security and governance challenges that hindered the full potential of using low-code tools in DevOps environments.

## 3 Methodology

We conducted semi-structured interviews to explore the integration of low-code platforms within DevOps environments. A total of twelve experienced IT professionals, working with both DevOps and low-code approaches. To recruit participants, invitations were initially sent to twenty-five IT professionals, and interviews were conducted with those who agreed to participate.

The interviews were conducted via Skype, Zoom, and Microsoft Teams, based on participant preference, and lasted approximately 40 minutes. Open-ended questions were used to explore participants' perspectives on merging DevOps with low-code approaches. All interviews were recorded and transcribed for subsequent analysis.

After data collection, we adopted Grounded Theory (GT) approach to extract themes and concepts for the interview transcripts [23]. The detailed steps followed for grounded theory are shown in Fig1.

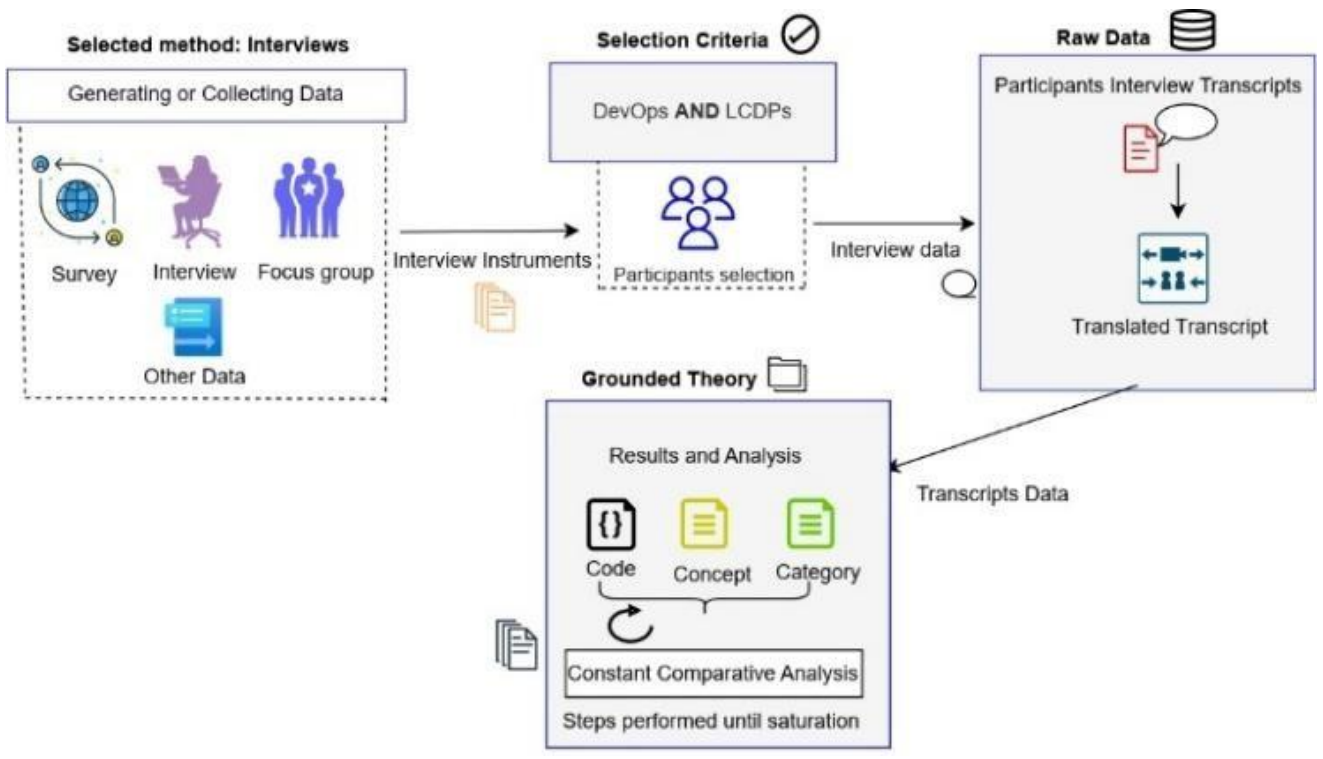

**Figure 1: Used research approach**

## 4 Results

The results and analysis are presented in this section. participants demographics details are presented in sub-section A. Emergence Categories are discussed in sub-section B, and sub-section C consists of security and governance practices and holistic framework.

### 4.1 Demographics

The demographic details of participants are presented in Table 1. Total twelve participants agreed to be a part of our study experiment. These participants were from three countries: Finland, Spain and China. Participants reported various roles (IT manager, Project manager, System analyst, Operational manager, and Developer). They also reported more than four years of experience in related domain.

**Table 1: Demographic details**

| Id | Role | Exp. | Domain | Tools | Country |
|---|---|---|---|---|---|
| 1 | ITm | 5 | Telecom | Github, App maker | Finland |
| 2 | Dev | >4 | Telecom | Github, App maker | Spain |
| 3 | PM | +7 | Software | Puppet, App maker | Spain |
| 4 | OPm | 5 | Software | Puppet, App maker | China |
| 5 | PM | 10 | Software | Chief, App maker, Appian | Finland |
| 6 | OPm | >5 | Software | Puppet, PowerApps | China |
| 7 | SA | +6 | Banking | Jenkins | Finland |
| 8 | SA | 7 | Banking | Jenkins, Gradle | China |
| 9 | SA | 5 | Software | Chief, Puppet, PowerApps | Spain |
| 10 | ITm | 4 | Software | Puppet, PowerApps | Spain |
| 11 | Dev | +6 | Telecom | Jenkins, TestOps | Finland |
| 12 | PM | +10 | Telecom | Jenkins | Spain |

Exp: Years of experience; PM: Project manager; SA: System analyst; Dev: Developer; ITm: IT manager; OPm: Operational manager; Software: Software development;

### 4.2 Emergence Categories

DevOps practitioners are seeing the impact of LCDPs on secure DevOps development process (e.g., "*in hybrid environment low-code will help in creating applications that automate your DevOps process*", id 4: OPm). When there is a large team working on application development, automation, and delivery, the need for security and governance becomes even more crucial (e.g., "*if you are working for large organization security and governance even become more important*", id 6: OPm).

Low-code and DevOps approaches are a mixed environment that presents vulnerabilities and are susceptible to threats and lack of visibility, just like any other (e.g., "*hybrid environments also have security vulnerabilities e.g., loss of resources and control, poor configuration, assess controls and even lack of visibility in terms of looking into resources*", id 4: OPm). Low-code platforms are also available for designing and implementing cybersecurity applications while utilizing a unified knowledge source (e.g., "*to opt anything with continuous environment is always challenging and should be explored to control security vulnerabilities*", id 3: PM)

Multiple cyber-attacks have focused our collective attention on securing our supply chains as shown by the Russian SolarWinds attacks by NOBELIUM in 2020 (e.g., "*there was Solarwinds hack with a high heap in cloud-based giants*, id 2: Dev). These recent events have highlighted the growing importance of keeping all deployed systems up to date with security updates (e.g., "*to make the process more secure, we need automated security scan and reliable tools or report generators to help enterprise to trust on maturity of using these platforms*, id 7: SA).

Attackers are increasingly seeking to exploit the way organisations work such as the use of unauthorised devices, applications, and infrastructure (e.g., "*the use of unauthorised tools also sometimes cause serious damage to the enterprise well*, id 5: PM). Even if LCDPs receive security upgrades on regular basis, this may not be enough to protect business applications from cyberattacks (e.g., *mitigating with DevOps can help with some issues but not all, however, well monitoring tools can identify vulnerabilities on time"*, id 2: Dev).

In addition, software practitioners are becoming more aware of vulnerabilities ("*low-code in DevOps is a double-edge sword*", id 11: Dev). In fact, proper cyber hygiene and implementation of basic security measures are often the best way to disrupt, prevent, and detect attacks (e.g., "*there are various loopholes…. but prevention is best…if you want to know solutions*", id 3: PM). However, risk must be considered as a whole and across the organisation since attackers are becoming more sophisticated (e.g., "*even after measuring security checks and audits, we cannot claim to secure 100 % of the application*, id 12: PM).

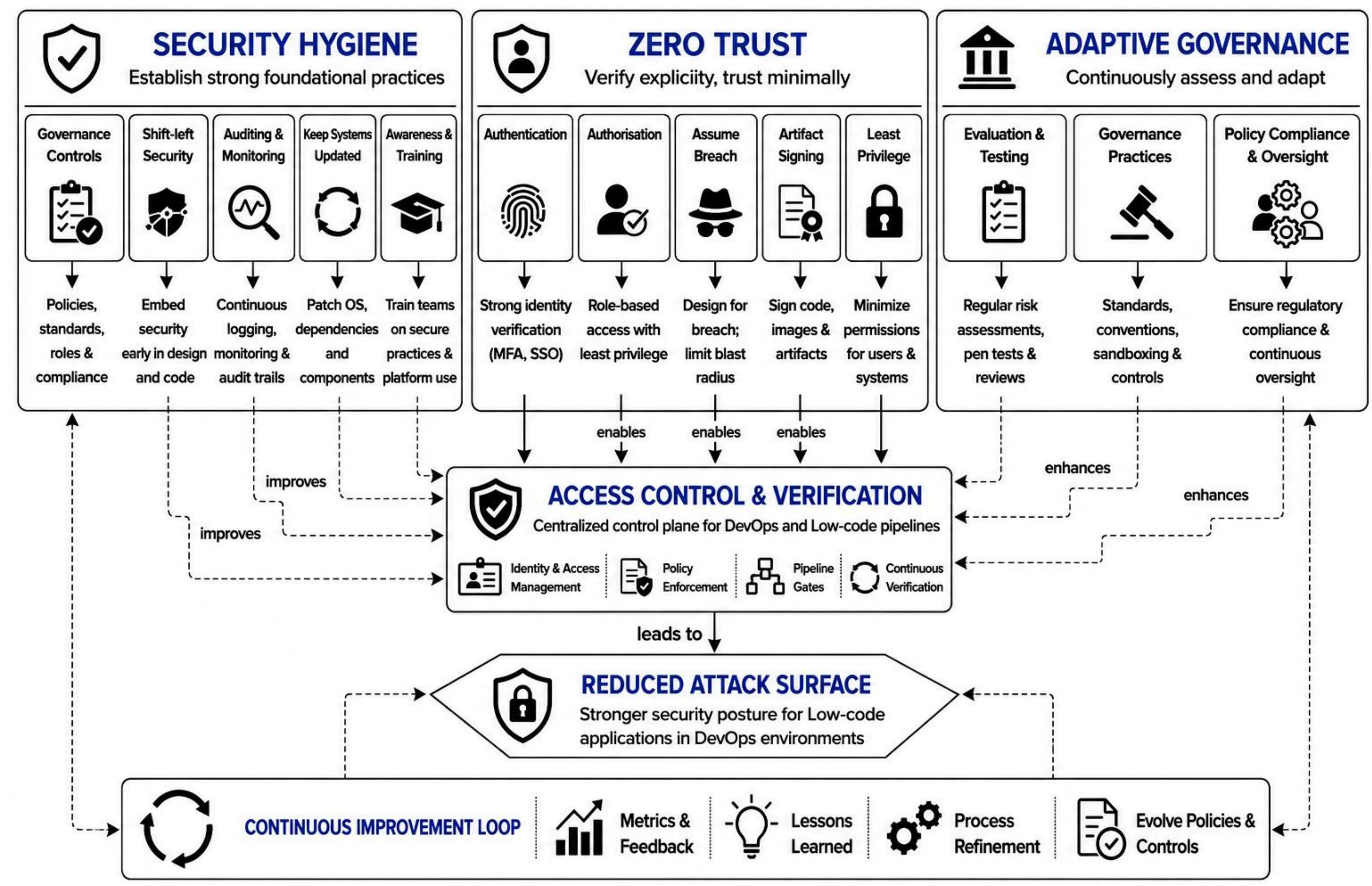

**Figure 2: Holistic framework Security and Governance Practice**

Moreover, although such a hybrid approach facilitates teamwork, security and data governance issues may not be top-of-mind for developers (e.g., "*They [LCDP] facilitate the development team but have issues like data security and governance*, id 2: Dev).

The effort for the integration of LCDPs into DevOps was also mentioned ("*it [such a hybrid environment] is not an easy migration*, developer) and raised the following open question: *do you think security and governance will be their top priority?*"

## 4.3 Recommended Security and Governance Practices

Based on an analysis of expert opinions, a holistic framework was developed (Fig 2) that illustrates the integration of security hygiene, Zero Trust principles, and adaptive governance to reduce the attack surface in DevOps environments.

### 4.3.1 Security Hygiene and DevOps Governance

- *Automated governance controls.* DevOps governance may be managed by considering integrating automated controls to secure the pipeline, such as least-privilege enforcement, automated policy checks, and CI/CD security gates [24].
- *Shift-left security.* Teams may consider adopting shift-left practices for security by embedding vulnerability scans and automated security testing earlier in CI/CD pipelines. This will help to identify issues before the deployment process [28].
- *Auditing.* Organisations can benefit from using monitoring and audit mechanisms to improve visibility over infrastructure, credentials, and user activities, supporting accountability and early threat detection [29].

### 4.3.2 Zero Trust Principles in the DevOps environment

- *Authentication.* DevOps teams can apply Zero Trust principles to ensure every access request is verified with least privilege, and systems are designed with breach-preparedness in mind [30]. Cloud-native environments can also be secured using continuous monitoring and verification at runtime [31].
- *Artifact signing & Policy as code.* DevOps pipelines can be secured by artifact signing and policy-as-code,extending Zero Trust practices into build and deployment workflows [32].

### 4.3.3 Adaptive governance in decentralised low-code environments

- *Evaluation and Testing.* Teams may apply vendor assessments, penetration testing, and standard security reviews, treating low-code applications the same as traditional ones [33].
- *Governance platforms for low-code development.* Developers can consider using governance practices such as consistent naming conventions, sandbox environments, and controlled release processes to keep low-code projects aligned with organisational policies [34]. As low-code decentralises development, organisations could also adopt frameworks that embed compliance, accountability, and oversight across distributed scenarios.

## 5 Discussion and Future Work

As LCDPs can be used as an attack vector, we need to consider the attack surface of DevOps environments. Given that it is hard to predict how technology will be exploited, we must assume that anything we develop or use could be a potential target.

LCDPs are designed to connect many tools which allow us fast-track integrations, e.g., streamline migrations. However, Low-code applications use many integrations with other services by means of APIs. Thus, these integrations should not only be continuously tested, as mentioned by [24], but also end-to-end supply chain security should be discussed. The likelihood of risk materialising also increases due to the huge acceleration in remote services, both at home and in the workplace, and the different levels of technical expertise between citizen and professional developers.

LCDPs can allow us to build custom mobile, web, and desktop apps that can satisfy advanced use cases. For complex automation use cases in which software professionals must develop APIs, customize front-end interfaces, or integrate the application with other systems, a more comprehensive platform that provides both low-code and coding capability can help manage and minimise security and governance challenges. However, there is a vendor lock-in that confines the existing features and cost.

Not all LCDPs offer features to deal with shadow IT [26] —all software or hardware used by employees is not approved by IT operations—, so IT experts must select one that adheres to their organisation's guidelines. Moreover, although a recent study [25] about testing facilities embedded in five well-known commercial LCDPs identified that security is one of the verification support features, LCDPs may not support all security features that are available in traditional coding languages.

Organisations are now facing an industrialised attacker economy with illegal commodity trading in which threat actors are skilled and relentless. To respond effectively, they can adopt governance and security practices such as automated governance controls, shift-left security, continuous auditing, explicit verification, artifact signing, policy-as-code, rigorous code testing, and the use of sandbox environments to keep low-code development platforms (LCDPs) aligned with security and governance (Section V). This approach seems viable in DevOps environments for creating a more secure approach that integrates LCDPs and might address the identified challenges by DevOps practitioners. When properly implemented, it implies a cultural shift that can enable us to unlock the potential of LCDPs. Therefore, it is also crucial we cultivate a security-conscious culture in addition to detection and protection. Beyond that, establishing governance requires a great deal of effort too, especially to set up tracking and alerting mechanisms and compliance measures. Thus, IT alignment at the operational level may be needed in this environment in addition to strategic alignment [4]. Indeed, DevOps experts [27] highlight that DevOps governance is essential to enable developers velocity.

# 6 Acknowledgements

The authors acknowledge the use of Grammarly and ChatGPT as supporting tools for language enhancement and manuscript refinement during the preparation of this paper. The authors have thoroughly reviewed, edited, and validated the final content and accept full responsibility for its accuracy and integrity.